\documentclass[aps,prb,preprint,groupedaddress,citeautoscript]{revtex4}

\usepackage{graphicx}
\usepackage{subfigure}
\usepackage{amsmath}
\usepackage{hyperref}
\usepackage{enumerate}

\newcommand{\point}[1]{\ensuremath{{\rm #1}}}
\newcommand{\subP}{_\point{P}}
\newcommand{\ie}{{\em i.e.,\ }}
\newcommand{\eg}{{\em e.g.,\ }}
\newcommand{\Ref}[2][.]{Ref#1~[\onlinecite{#2}]}
\newcommand{\Sec}[2][Section]{#1~\ref{sec:#2}}
\renewcommand{\vec}[1]{\mathbf{#1}}
\newcommand{\vecsym}[1]{\boldsymbol{#1}}

\newcommand{\eqnref}[1]{\eqref{eq:#1}}
\newcommand{\Eq}[1]{Eq.~\eqnref{#1}}
\newcommand{\Equation}[1]{Equation~\eqnref{#1}}

\begin{document}

\title{Inadequacies in the conventional treatment
of the radiation field of moving sources}

\author{Houshang Ardavan}
\affiliation{Institute of Astronomy, University of Cambridge,
Madingley Road, Cambridge CB3 0HA, UK}
\author{Arzhang Ardavan}
\affiliation{Clarendon Laboratory, Department of Physics, University of Oxford,
Parks Road, Oxford OX1 3PU, UK}
\author{John Singleton}
\affiliation{National High Magnetic Field Laboratory, MS-E536,
Los Alamos National Laboratory, Los Alamos, New Mexico 87545}
\author{Joseph Fasel}
\author{Andrea Schmidt}
\affiliation{Process Modeling and Analysis, MS-E548,
Los Alamos National Laboratory, Los Alamos, New Mexico 87545}

\date{2009 May 28}

\begin{abstract}
There is a fundamental difference
between the classical expression for the retarded electromagnetic potential
and the corresponding retarded solution
of the wave equation that governs the electromagnetic field.
While the boundary contribution
to the retarded solution for the {\em potential}
can always be rendered equal to zero
by means of a gauge transformation
that preserves the Lorenz condition,
the boundary contribution
to the retarded solution of the wave equation governing the {\em field}
may be neglected only if it diminishes with distance
faster than the contribution of the source density in the far zone.
In the case of a source whose distribution pattern both rotates
and travels faster than light {\em in vacuo},
as realized in recent experiments,
the boundary term in the retarded solution governing the field
is by a factor of the order of $R^{1/2}$ {\em larger}
than the source term of this solution
in the limit that the distance $R$ of the
boundary from the source tends to infinity.
This result is consistent
with the prediction of the retarded potential
that part of the radiation field generated
by a rotating superluminal source
decays as $R^{-1/2}$, instead of $R^{-1}$,
a prediction that is confirmed experimentally.
More importantly,
it pinpoints the reason why an argument
based on a solution of the wave equation governing the field
in which the boundary term is neglected
(such as appears in the published literature)
misses the nonspherical decay of the field.
\end{abstract}
\maketitle

\section{Introduction\label{sec:intro}}

Scientific investigation of the electromagnetic field
generated by a charged particle that moves faster than light
began with a largely ignored article
by physicist and mathematician Oliver Heaviside in 1887 \cite{Heaviside:1894}
and is the subject of several papers by Sommerfeld in 1904 and early 1905
\cite{Sommerfeld:1904}.
The publication of the special theory of relativity in June 1905 \cite{Einstein:1905}, however,
discouraged further work as one of its tenets is, of course,
that any known particle that has a charge also has a rest mass
and so is barred from moving faster than light.
Moreover,
no source that moves faster than the wave speed
can be pointlike,
for this results in infinitely strong fields
on the envelope of the emitted wave fronts.
It was not until the early 1970s
that Bolotovskii and Ginzburg \cite{Bolotovskii:1972,Ginzburg:1972}
pointed out that although special relativity
precludes massive particles from moving superluminally,
patterns of distribution of extended sources can move faster than light
as a result of the coordinated motion of their constituent charged particles,
which themselves remain subluminal
\cite{ArdavanA:2004,Singleton:2004,ArdavanH:2004,ArdavanH:2007,Schmidt:2007}.
The charge separation resulting from such coordinated motion
gives rise to a polarization current
whose distribution pattern can move faster than light {\em in vacuo.}
The way in which this current acts as a source of radiation
is immediately apparent from the  Amp\`ere-Maxwell equation (SI units),
\begin{equation} \label{eq:M4}
\vecsym{\nabla\times}\vec{H} = \vec{J}+\frac{\partial\vec{D}}{\partial t}
= \vec{J}+\epsilon_0\frac{\partial\vec{E}}{\partial t}+\frac{\partial\vec{P}}{\partial t},
\end{equation}
in which we see that the polarization current $\partial\vec{P}/{\partial t}$
contributes to the magnetic field $\vec{H}$
in just the same way as the current $\vec{J}$ of free charges.
However,
as $\partial\vec{P}/\partial t$ is not carried by massive particles,
it is not limited to subluminal speeds.
We should clarify that while the source distribution is superluminal,
the emitted radiation (as any other)
travels {\em at} the speed of light.

Extended sources of electromagnetic radiation
whose distribution patterns move faster than light {\it in vacuo}
have been experimentally realized by several groups
\cite{Bolotovskii:2005,ArdavanA:2004,Singleton:2004,Besarab:2004}
and have many potential applications in science and technology \cite{ArdavanA:1999}.
Furthermore,
they may be responsible for the extreme properties
of the electromagnetic radiation received
from astronomical objects such as pulsars
(rapidly spinning, highly magnetized neutron stars)
\cite{ArdavanH:2008,Schmidt:2007,ArdavanH:2009}.
Methods of dealing with the asymptotic behavior
of the radiation emitted by such sources, however,
are largely neglected in standard texts \cite{Jackson:1999}.
More dangerously,
certain textbook formul\ae,
derived in the context of stationary or subluminally moving sources,
cannot be used in treating unusual cases
such as those in which the radiation remains focused in the far zone
\cite{ArdavanH:2004,ArdavanH:2007,ArdavanH:2008a}.
An example is this expression
for the far-zone magnetic field of a localized current with the density $\vec{j}$:
\begin{equation}
\vec{B}(\vec{x}\subP,t\subP)\simeq\frac{1}{c}\int{\rm d}^3x
\frac{\left[\vecsym{\nabla\times}\vec{j}\right]}{\left|\vec{x}\subP-\vec{x}\right|}.
\label{eq:2}
\end{equation}
Here, $(\vec{x}\subP,t\subP)$ and $(\vec{x},t)$
are the space-time coordinates of the observation point and the source points, respectively,
and the square brackets denote the retarded value of $\vecsym{\nabla\times}\vec{j}$.
(Note that,
in contrast to the free-charge current $\vec{J}$ of \Eq{M4},
this $\vec{j}$ is a generalized current
that also includes $\partial\vec{D}/\partial t$.)
The derivation of \Eq{2}
involves the neglect, in the far zone,
of a boundary term containing the gradient of the radiation field.
In the case of a conventional source,
this term decays more rapidly with distance
than the integral that remains in \Eq{2}.
This is not the case, however,
for the radiation field of a polarization current
whose distribution pattern has an accelerated motion
with a speed exceeding that of light, $c$
\cite{ArdavanH:2008a,ArdavanH:2000,ArdavanH:1999}.
In the next section,
we describe such a source,
the emission from which consists of a collection of narrowing subbeams
for which the absolute value of the gradient of the radiation field $\vec{B}$
{\em increases} (as $R\subP^{7/2}$) with the distance $R\subP$ from its source.

Since the exact form of \Eq{2}
entails an additional boundary term
that depends on the gradient of the field in the far zone
[\Eq{9}, below],
the increase in the magnitude of the gradient
of the present radiation field with distance
renders the boundary term in the retarded solution of Maxwell's equation
for $\vec{B}$ dominant over its source term.
The inadequacy of \Eq{2}
in describing the radiation field generated by a superluminal source
lies in the neglect of this boundary term.

The remainder of this paper is organized as follows.
Section~II describes the superluminally-rotating source distribution,
and a practical experimental realization of such a source.
In order to make the
later discussion more comprehensible, Section III considers a small volume
element of this source and discusses the superposition of multiple
retarded times that render the field of a point source divergent. 
Section~IV generalizes this
discussion to an extended source and describes the morphology
of the resulting radiation ``beam'', providing a geometrical argument
as to why the boundary term missing from Eq.~2 dominates in the
far field, and Section~V describes supporting experimental data. 
Sections VI and VII contain the
main, substantive point of the current work;
while the boundary contribution to the retarded solution for the
potential can always be rendered equal to zero by means of a gauge
transformation, the boundary contribution to the retarded solution of the
wave equation governing the field cannot be neglected in the case of a
superluminal source. Finally Section VIII gives a
brief summary.

\section{A superluminally rotating source distribution\label{sec:source}}

Here we consider a particular moving source distribution
for which \Eq{2} is not an adequate description
of the radiation field in the far zone and its experimental realization.
The findings of \Ref[s.]{ArdavanH:2007,ArdavanH:2008a}
fully clarify how this case should be treated
in order to obtain results consistent with experimental data.
In these papers,
we solved the inhomogeneous wave equation
governing the electromagnetic potential [\Eq{4} below]
in unbounded space,
under null initial conditions,
with a polarization current density $\vec{j}=\partial\vec{P}/\partial t$
for which
\begin{equation}
P_{r,\varphi,z}(r,\varphi,z,t)=
s_{r,\varphi,z}(r,z)\cos(m{\hat\varphi})\cos(\Omega t),
\qquad -\pi<{\hat\varphi}\le\pi,
\label{eq:-1}
\end{equation}
where
\begin{equation}
{\hat\varphi}\equiv\varphi-\omega t.
\label{eq:0}
\end{equation}
Here, $P_{r,\varphi,z}$ are the components of the polarization $\vec{P}$
in a cylindrical coordinate system based on the axis of rotation,
$\omega$ and $\Omega$ are angular frequencies,
$\vec{s}(r,z)$ is an arbitrary vector function with a finite
support in $r>c/\omega$,
and $m$ is a positive integer.
For fixed $t$,
the azimuthal dependence of the polarization (\ref{eq:-1})
along each circle of radius $r$ within the source
is the same as that of a sinusoidal wave train,
of wavelength $2\pi r/m$,
whose $m$ cycles fit around the circumference of the circle smoothly.
As time elapses,
this wave train both propagates around each circle
with the superluminal velocity $r\omega$
and oscillates in its amplitude
with frequency $\Omega$.
This is a generic source:
one can construct any distribution
with a uniformly rotating pattern,
$P_{r,\varphi,z}(r,{\hat\varphi},z)$,
by the superposition over $m$
of terms of the form $s_{r,\varphi,z}(r,z,m)\cos(m{\hat\varphi})$.
For example, the polarization-current patterns responsible for
the pulsar radiation visible from Earth are thought to be animated by
the superluminal rotation of the neutron-star core's intense magnetic
field through the pulsar's plasma atmosphere~\cite{ArdavanH:2008}; one can
envisage representing this rotating pattern by summing such terms.
Moreover, Eq.~(3)~ corresponds to a laboratory-based source
that has been used in experimental demonstrations
of some of the phenomena described below
\cite{ArdavanA:2004,Singleton:2004}.

\begin{figure}
\centering
\subfigure{\includegraphics[height=7cm]{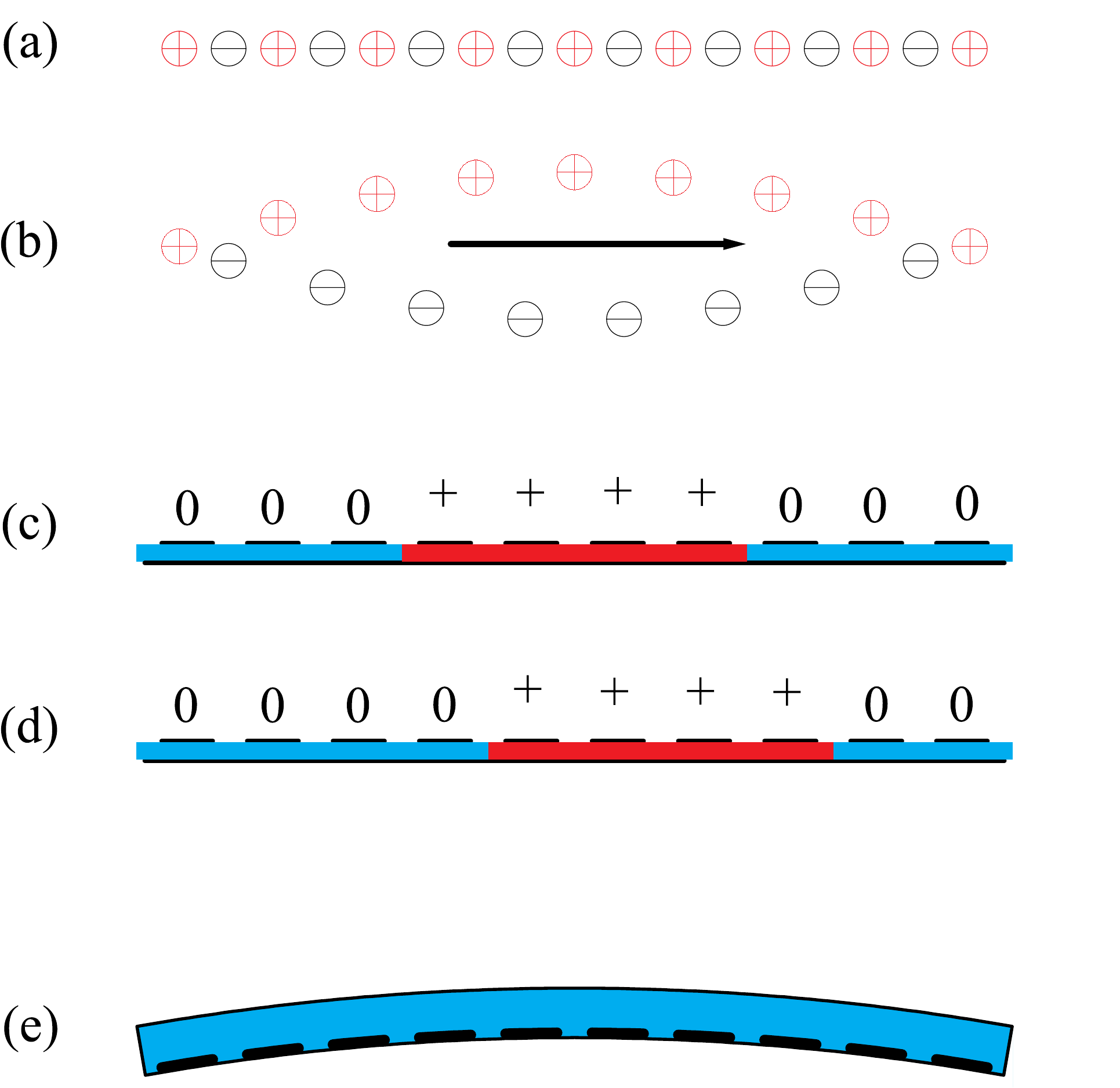}}
\subfigure{\includegraphics[height=7cm]{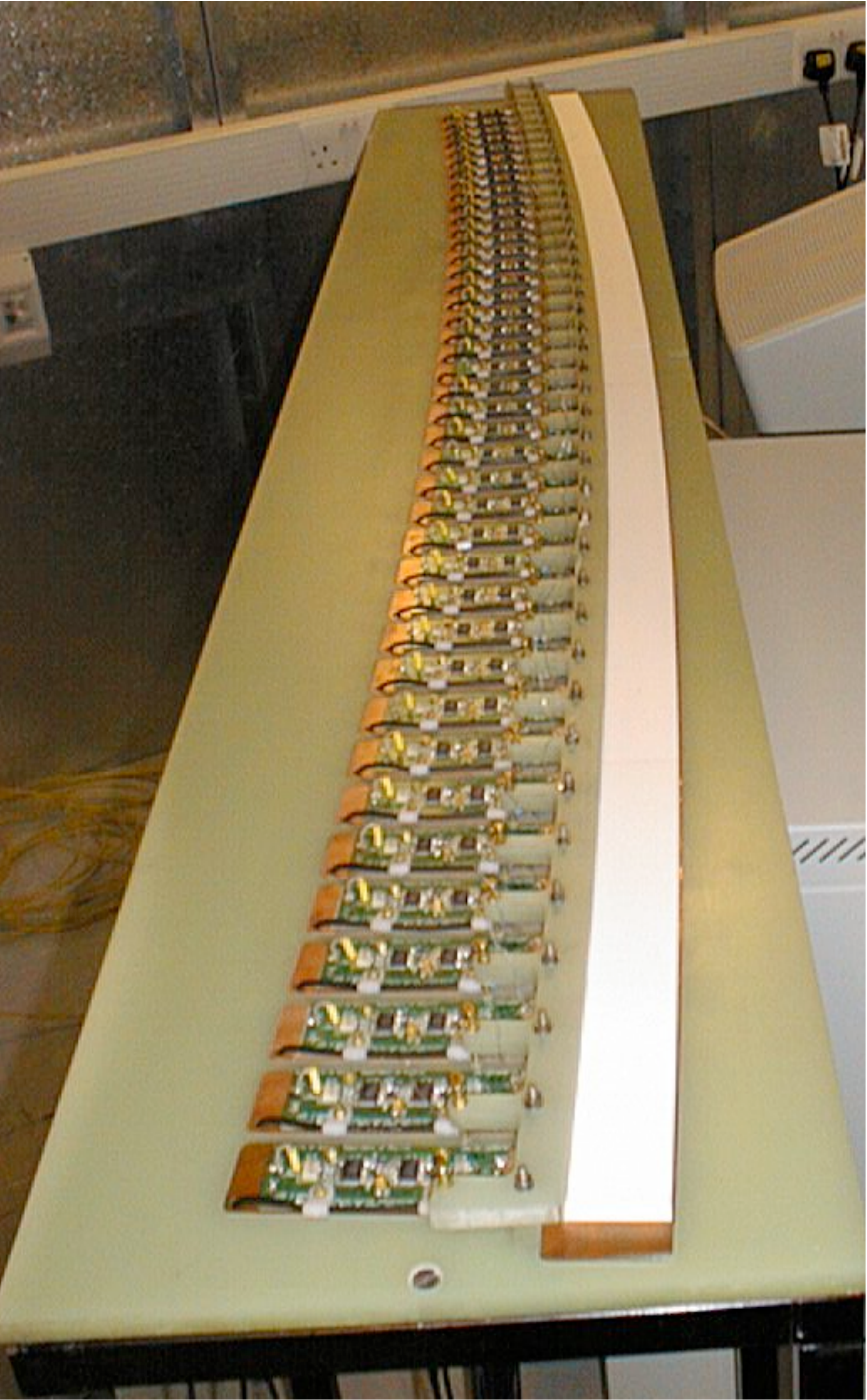}}
\caption{Left: experimental animation of a superluminal polarization
current.  (a)~A simplified dielectric solid containing negative
($\ominus$) and positive ($\oplus$) ions. In (b), a spatially-varying
electric field has been applied, causing the positive and negative
ions to move in opposite directions; a finite polarization {\bf P} has
therefore been induced.  If the spatially-varying field is made to
move along the direction of the arrow, the polarized region moves
with it.  (c)~Schematic side view of a practical superluminal source,
showing metal electrodes above a strip of dielectric (shaded region)
and a ground plate below it.  ``0'' indicates that there is no voltage
on that particular upper electrode; the symbol + indicates a positive
voltage applied to the upper electrode.  The voltage on the electrodes
produces a finite polarization of the dielectric (darker shading).
(d)~By switching the voltages on the electrodes on and off, the
polarized region (darker shading) can be made to move along the
dielectric.  (e)~Top view, showing the curvature of the dielectric
(lighter shaded region). The curvature 
introduces centripetal acceleration in
the moving polarized region.  Note that the electrodes (black
shading) cover only part of the top surface of the dielectric.
Right: a photograph of the experimental 
apparatus showing a curved strip of dielectric material (alumina)
sandwiched between a copper ground plate (below) and 41 electrodes,
each connected to its own amplifier (left).} 
\label{fig0}
\end{figure}

The experimental apparatus in question (see Fig.~\ref{fig0}) 
consists of a 
continuous strip of dielectric material on top of 
which is placed an array of metal electrodes; 
underneath is a continuous ground plate.
Each upper electrode is connected to an individual 
amplifier, so that by turning the amplifiers on and off 
in sequence one can apply voltages that
generate a polarized region that 
moves along the dielectric with an arbitrarily high speed.
In practice, the dielectric is a strip of alumina 10~mm thick 
and 50~mm across, corresponding to a 10$^\circ$ arc of 
a circle of average radius 10.025~m.
Above the alumina strip, there are 41 upper electrodes of 
mean width 42.6~mm, with centers 44.6~mm apart 
\cite{ArdavanA:2004,Singleton:2004}.

The voltage $V_j$ applied to the $j$th electrode 
is of the form
\begin{equation} 
V_j=V_0\cos[m\omega(j\Delta t-t)]\cos(\Omega t),
\label{e13}
\end{equation}
where $m\omega$ and $\Omega$ are as defined in \Eq{-1}:
the first cosine gives rise to the polarization-current 
wave that propagates along the dielectric and the 
second to a modulation of this wave.
The source speed $v$ is determined by the phase 
difference between the oscillations of neighboring 
electrodes; given the dimensions of the electrodes 
of the experimental machine,
a superluminal speed $v\geq c$ is obtained for 
$\Delta t\leq 148.8$~ps.  In the experiments reported 
in \Ref[s.]{ArdavanA:2004,Singleton:2004}, 
$m\omega/(2\pi)$ and $\Omega/(2\pi)$ 
had the values 552.645 MHz and 46.042 MHz, 
respectively, and the source was run at $v/c=0.875$,
1.064, 1.25 and 2.00 (see \Ref[s.]{ArdavanA:2004,Singleton:2004}
for a more complete description).

\section{The field generated by a constituent volume element of the 
source\label{sec:volumeelement}}

A superluminal source is necessarily volume-distributed
\cite{Bolotovskii:1972}.  However, its field can be built up 
from the superposition of the fields 
of its small, moving constituent volume elements.  
Figure~\ref{fig1}(a) shows that the waves generated by such a 
volume element of a rotating superluminal source 
possess a cusped envelope and that, inside the envelope, 
{\it three} wave fronts pass through any given observation 
point simultaneously.  This reflects the fact that the field 
inside the envelope receives simultaneous 
contributions from three distinct values of the retarded time 
[see Fig.~\ref{fig1}(c)].  On the cusp of the envelope, 
where the space-time trajectory of the source is tangent 
to the past light cone of the observer [Fig.~\ref{fig1}(d)], 
all three contributions toward the value of the field 
coalesce~\cite{ArdavanH:1998,ArdavanH:2003,ArdavanH:2004}.  

\begin{figure}
\centering
\includegraphics[height=4cm]{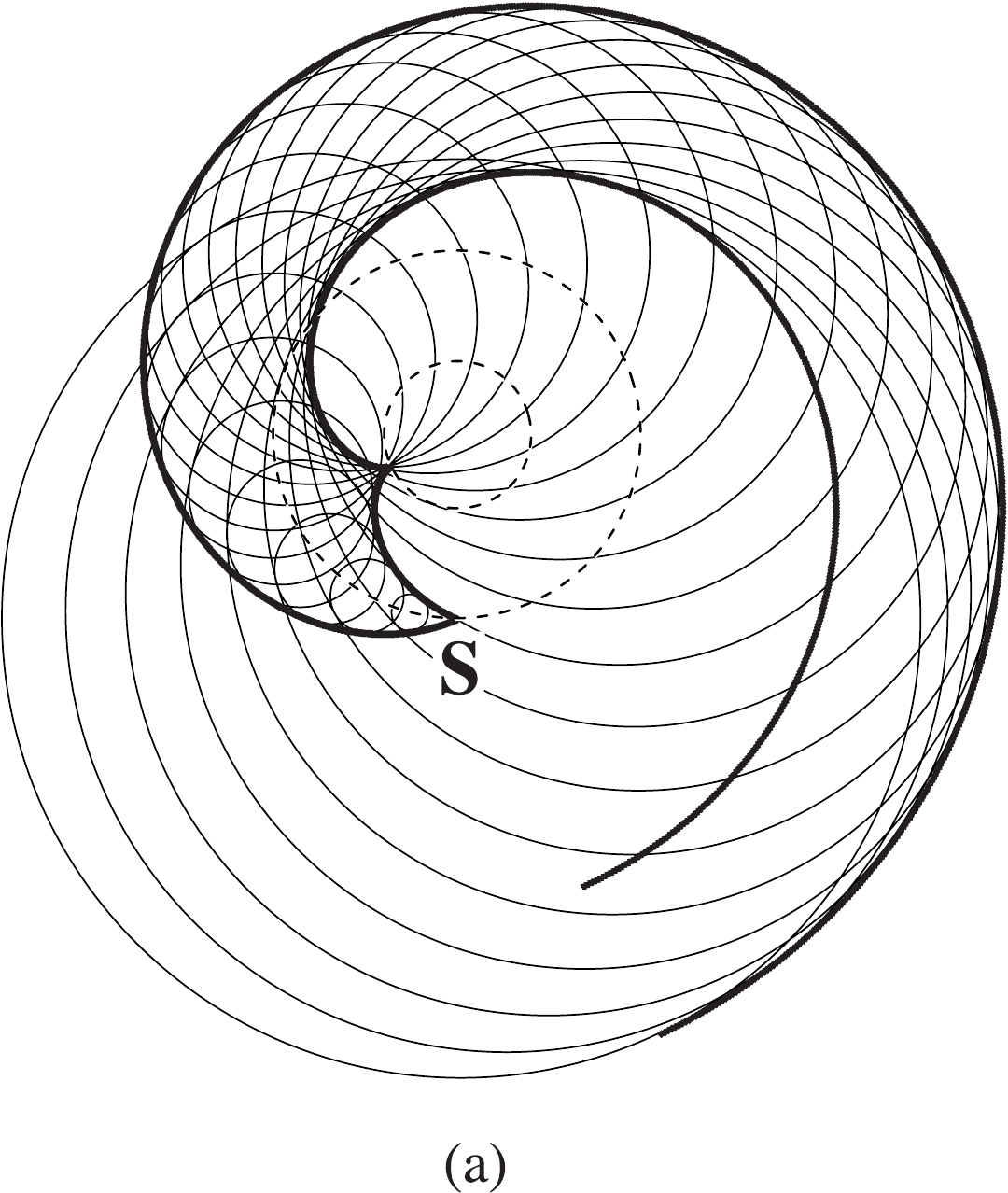} 
\end{figure} 

\begin{figure}
\centering
\subfigure{\includegraphics[height=3cm]{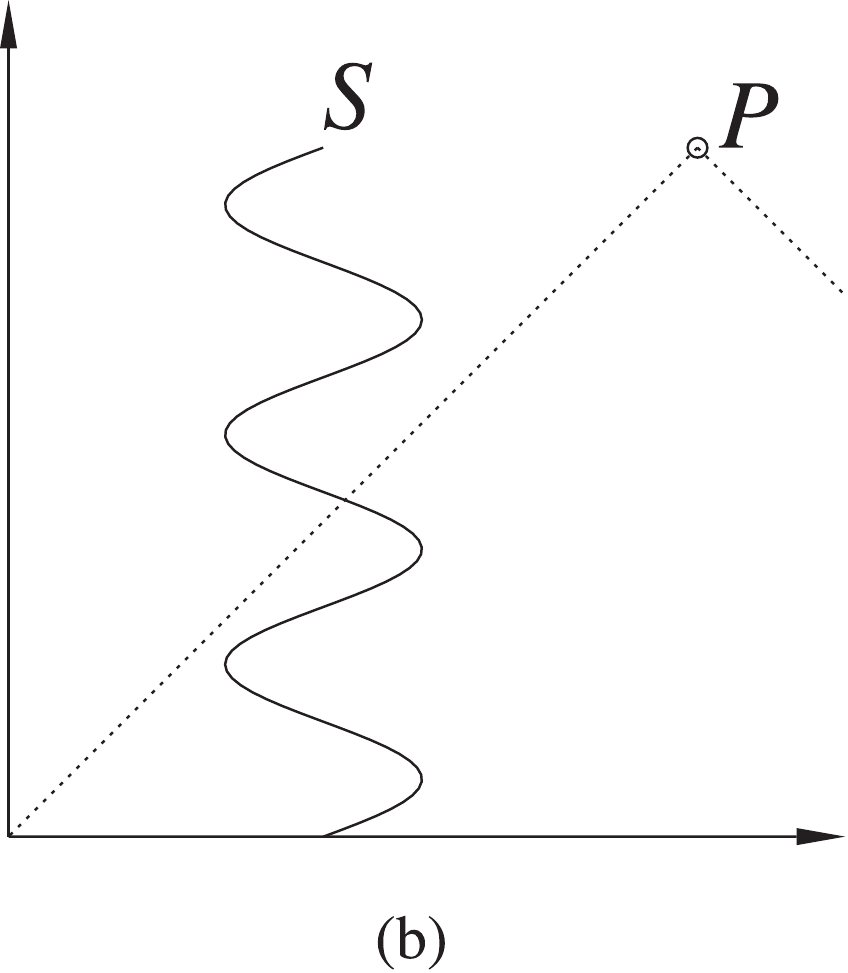}}
\subfigure{\includegraphics[height=3cm]{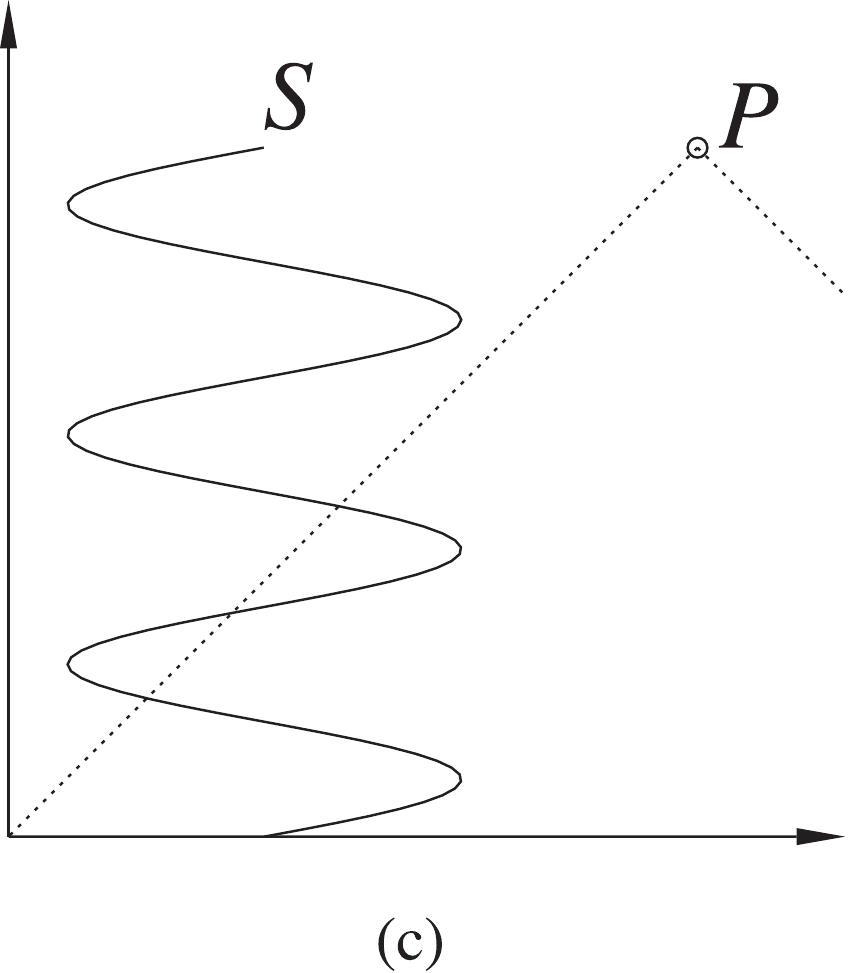}}
\subfigure{\includegraphics[height=3cm]{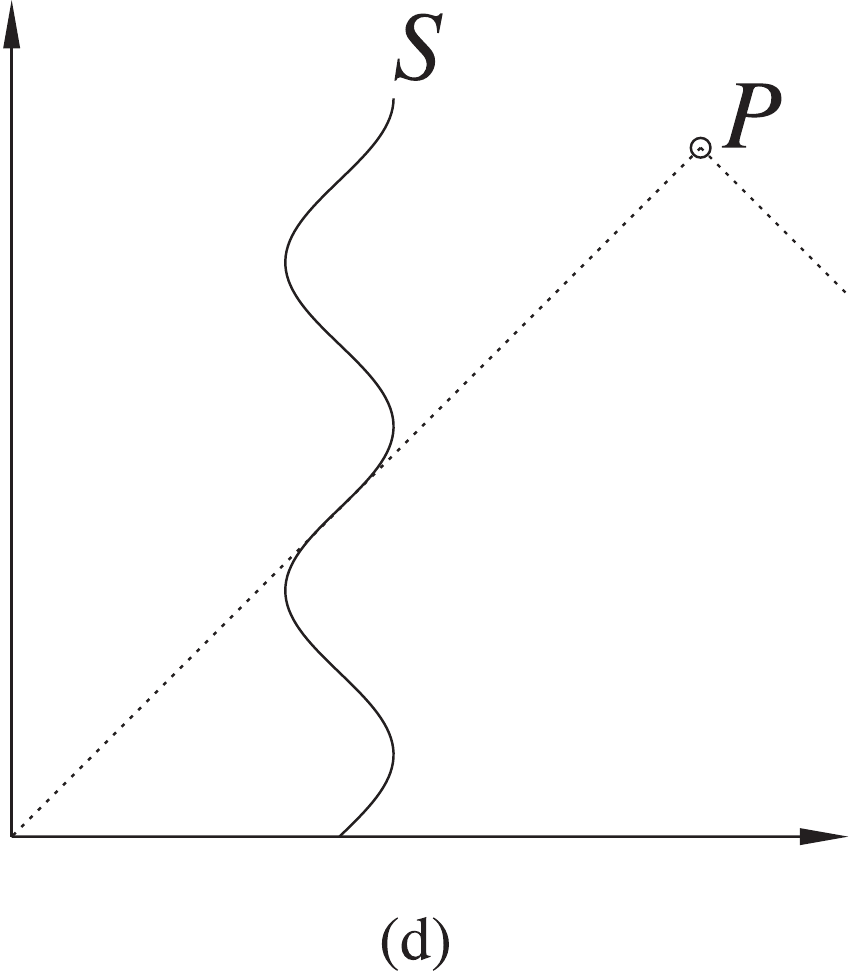}}
\caption{(a) Envelope of the spherical wave fronts 
emanating from a superluminally moving source 
element (S) in uniform circular motion. The lighter circles
are Huygen's wavelets emitted by the source
as it traverses its circular orbit, designated by the larger 
of the two dotted circles. The heavy curves 
show the cross section of the envelope of the wavelets
within the 
plane of the orbit of the source. The smaller of
the dotted circles represents the light cylinder $r=c/\omega$.  
(b), (c) and (d) Space-time ({\it i.e.} $ct$ versus 
distance $x$)
diagrams showing the intersection of the trajectory 
of the source point S with the past light cone of the 
observation point P when P lies outside (b), inside (c), 
and on the cusp of (d) the envelope of wave fronts.}
\label{fig1}
\end{figure} 

On this cusp (caustic), the source approaches the observer 
with the speed of light and zero acceleration at the retarded time, 
i.e.\ ${\rm d}R(t)/{\rm d}t=-c$ and ${\rm d}^2R(t)/{\rm d}t^2=0$, 
where $R(t)\equiv\vert{\bf x}(t)-{\bf x}\subP\vert$ 
is the distance between the source point ${\bf x}(t)$ 
and the observation point ${\bf x}\subP$. 
As a result, the interval of emission time 
for the signal carried by the cusp is 
much longer than the interval of its reception time~\cite{ArdavanH:2003}.
Fig.~\ref{fig1}(d) represents this diagrammatically;
rather than merely intersecting at one or three points,
the space-time trajectory of the source and the observer's
light cone become tangent 
to one another.
This essentially instantaneous reception of contributions 
from an extended period of emission time represents focusing 
of the radiation in the time domain; it has been
described as {\it temporal focusing}
and is the subject of two experimental papers~\cite{ArdavanA:2004,Singleton:2004}.
It is this temporal focusing that leads to the
divergence of the field of a point source on the cusp of the envelope of
wave fronts and is ultimately responsible for the component of the
radiation from an extended source whose intensity 
decays as 1/distance,
rather than the inverse square law that
applies to all other sources~\cite{ArdavanH:2007,ArdavanH:2008a};
consequently, this part of the radiation will
dominate in observations made from large distances.

\begin{figure} 
\centering
\includegraphics[height=7cm]{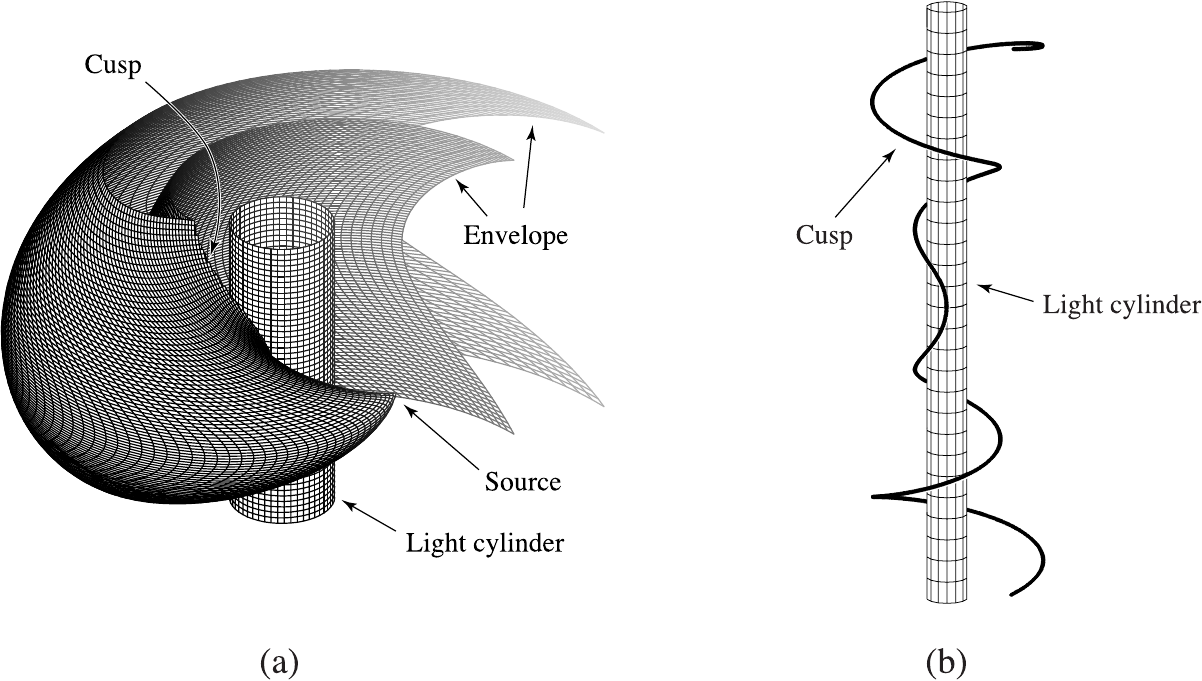}
\caption{Three-dimensional views of the envelope (a) and its cusp (b).}
\label{fig2}
\end{figure}

A three-dimensional view of the envelope of wave fronts 
and its cusp is shown in Fig.~\ref{fig2}.  
The two sheets of the envelope, and the cusp along 
which these two sheets meet tangentially, 
spiral outward into the far zone.  
In the far zone, the cusp lies on the double cone 
$\theta\subP=\arcsin[c/(r\omega)]$, $\theta\subP=\pi-\arcsin[c/(r\omega)]$, 
where $(R\subP, \theta\subP, \varphi\subP)$ denote the spherical 
polar coordinates of the observation point P.  
Thus, a stationary observer in the polar interval 
$\arcsin[c/(r\omega)]\le\theta\subP\le\pi-\arcsin[c/(r\omega)]$ 
receives recurring pulses from each volume element
as the envelope rotates 
past him/her~\cite{ArdavanH:2007}.

\begin{figure}
\centering
\includegraphics[height=8cm]{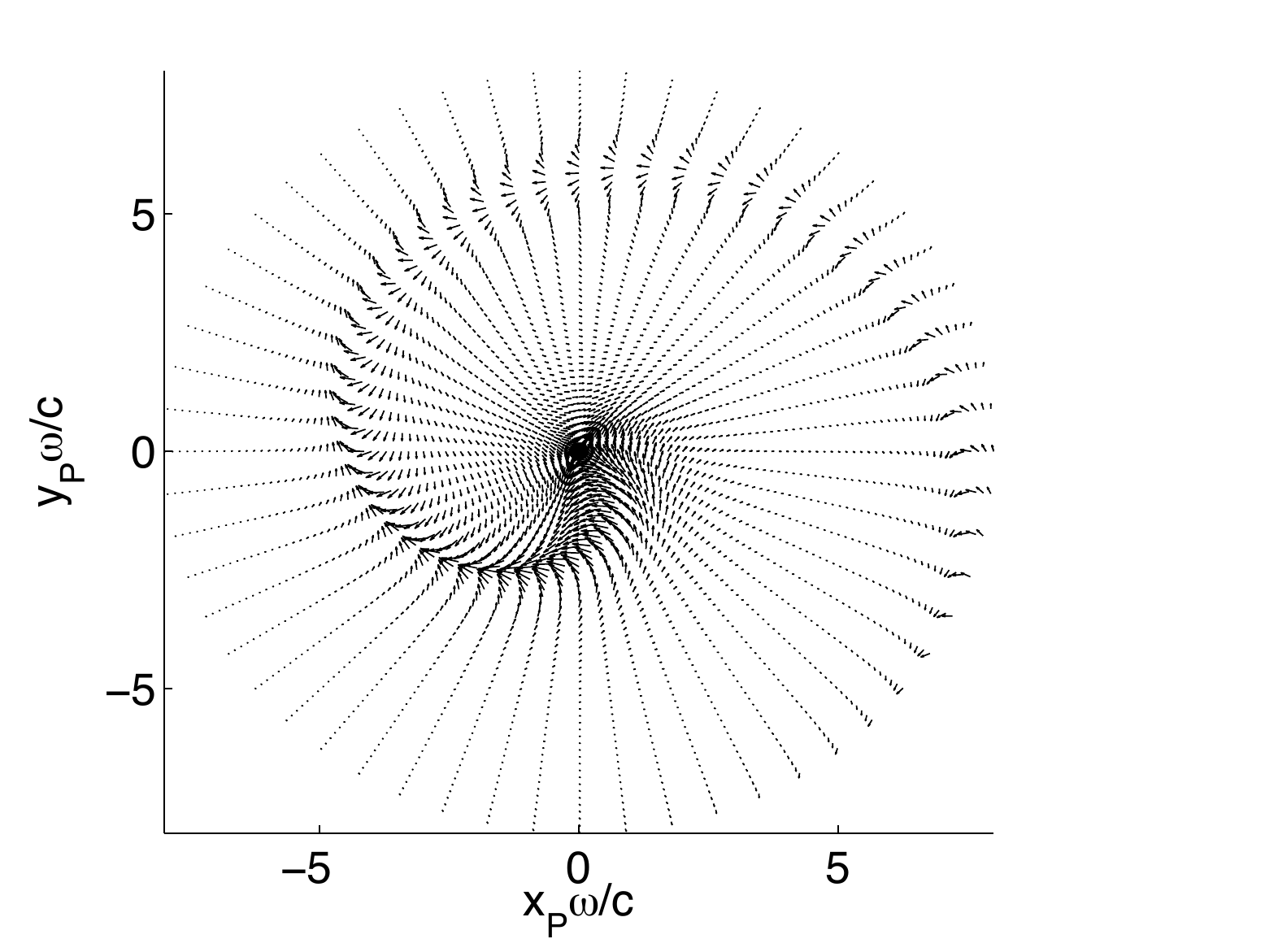}
\caption{Polarization position angles and field strengths 
on the cone $\theta\subP=\pi/12$ outside the envelope 
of wave fronts for a source element with $r\omega=2$.
Here, the field vectors are projected onto the plane $(x\subP,y\subP)$
of the source's orbit.}
\label{fig3}
\end{figure}

Figure~\ref{fig3} shows the radiation field generated 
by the rotating source element $S$ on a cone close to the cusp, 
just outside the envelope.  Not only does the spiraling 
cusp embody a recurring pulse, but the plane of polarization 
of the radiation swings across the pulse~\cite{Schmidt:2007},
as does the radio emission received from pulsars~\cite{Lyne:2006}.  

\section{The field generated by the entire volume of the source
\label{sec:volume}}

\begin{figure}
\centering
\includegraphics[height=7cm]{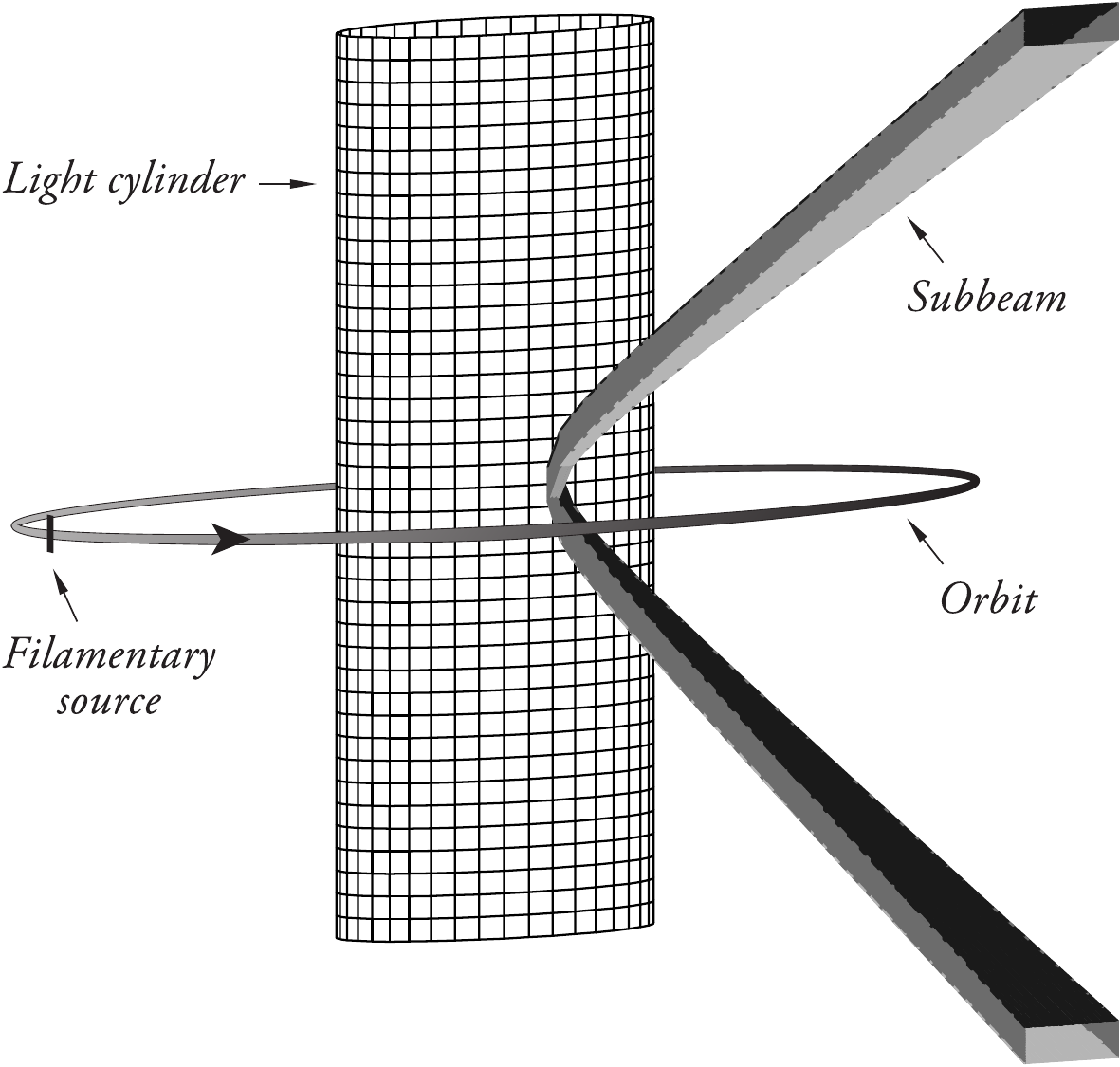}
\caption{Schematic illustration of the light cylinder $r=c/\omega$, 
the filamentary part of the source that approaches 
the observation point with the speed of light and 
zero acceleration at the retarded time, 
the orbit $r=c/(\omega\sin\theta\subP)$ of this filamentary source, 
and the subbeam formed by the bundle of cusps that emanate 
from the constituent volume elements of this filament.}
\label{fig6}
\end{figure}

The dominant contribution towards the field of an extended 
source comes from a thin filamentary part of the source that 
approaches the observer, along the radiation direction, 
with the speed of light and zero acceleration at the 
retarded time~\cite{ArdavanH:2007}.  For an observation point 
P in the far zone with the coordinates 
$(R\subP,\theta\subP,\varphi\subP)$, this filament is located at 
$r=(c/\omega)\csc\theta\subP$, $\varphi=\varphi\subP+3\pi/2$ 
and is essentially parallel to the rotation axis (Fig.~\ref{fig6}).  
The collection of cusps of the envelopes of wave fronts that 
emanate from various volume elements of the contributing filament 
form a subbeam whose polar width is nondiffracting: 
the linear dimension of this bundle of cusps in the 
direction parallel to the rotation axis remains the same 
at all distances from the source, so that 
the polar angle $\delta\theta\subP$ subtended by the subbeam 
decreases as ${R\subP}^{-1}$ with increasing $R\subP$ 
(Fig.~\ref{fig6} and~\Ref{ArdavanH:2007}).  

In that it consists of caustics and so is constantly 
dispersed and reconstructed out of other waves, 
the subbeam in question radically differs from 
a conventional radiation beam (see Appendix D of \Ref{ArdavanH:1998}).  
The narrowing of its polar width (as ${R\subP}^{-1}$) is accompanied 
by a more slowly diminishing intensity (an intensity that 
diminishes as ${R\subP}^{-1}$ instead of ${R\subP}^{-2}$ with distance), 
so that the flux of energy across its cross sectional area remains 
the same for all $R\subP$~\cite{ArdavanH:2007}. 
As we will see in the next section, this slower rate of decay 
of the emission from a superluminally rotating source has been 
confirmed experimentally upto several hundred Fresnel distances 
(Rayleigh ranges)~\cite{ArdavanA:2004,Singleton:2004}.

The contributing part of an extended source 
(the filament that approaches the observation 
point with the speed of light and zero acceleration) 
changes as the source rotates (see Fig.~\ref{fig6}).  
In the case of a turbulent plasma with a superluminally 
rotating macroscopic distribution, therefore, 
the overall beam within which the narrow, nonspherically 
decaying radiation is detectable would consist of an 
incoherent superposition of coherent, nondiffracting 
subbeams with widely differing amplitudes and phases 
(similar to the train of giant pulses received from the 
Crab pulsar \cite{Slowikowska:2005}). 

The overall beam occupies a solid angle whose polar 
and azimuthal extents are independent of the distance $R\subP$.  
It is detectable within the polar interval 
$\arccos[(r_l\omega/c)^{-1}]\le\vert\theta\subP-\pi/2\vert
\le\arccos[(r_u\omega/c)^{-1}]$, where $[r_l,r_u]$ denotes 
the radial extent of the superluminal part of the source. 
The azimuthal profile of this overall beam reflects the 
distribution of the source density around the cylinder 
$r=c/(\omega\sin\theta\subP)$, from which the dominant 
contribution to the radiation arises~\cite{ArdavanH:2007}. 

Because the subbeams that constitute the overall beam 
are narrower the farther away from the source they are 
detected, the absolute value of the gradient of the radiation 
field associated with them {\em increases} with distance: 
it has been explicitly
shown in \Ref{ArdavanH:2008a} that the gradient of the nonspherically decaying 
field generated by the superluminal source described in 
\Eq{-1} has a magnitude that increases as 
${R\subP}^{7/2}$ with increasing $R\subP$.
\section{Experimental and observational confirmation that the nonspherically-decaying term dominates
in the far field}
The issue under discussion in this paper is the dominance of the boundary
term neglected in Eq.~(2) in the far field.
Experimental demonstrations of the far-field
dominance of the boundary term focus on detecting
emissions whose intensity decays with distance
as $1/R_{\rm P}$, rather that the more conventional inverse-square law.

\begin{figure}[tbp]
   \centering
\includegraphics[height=9cm]{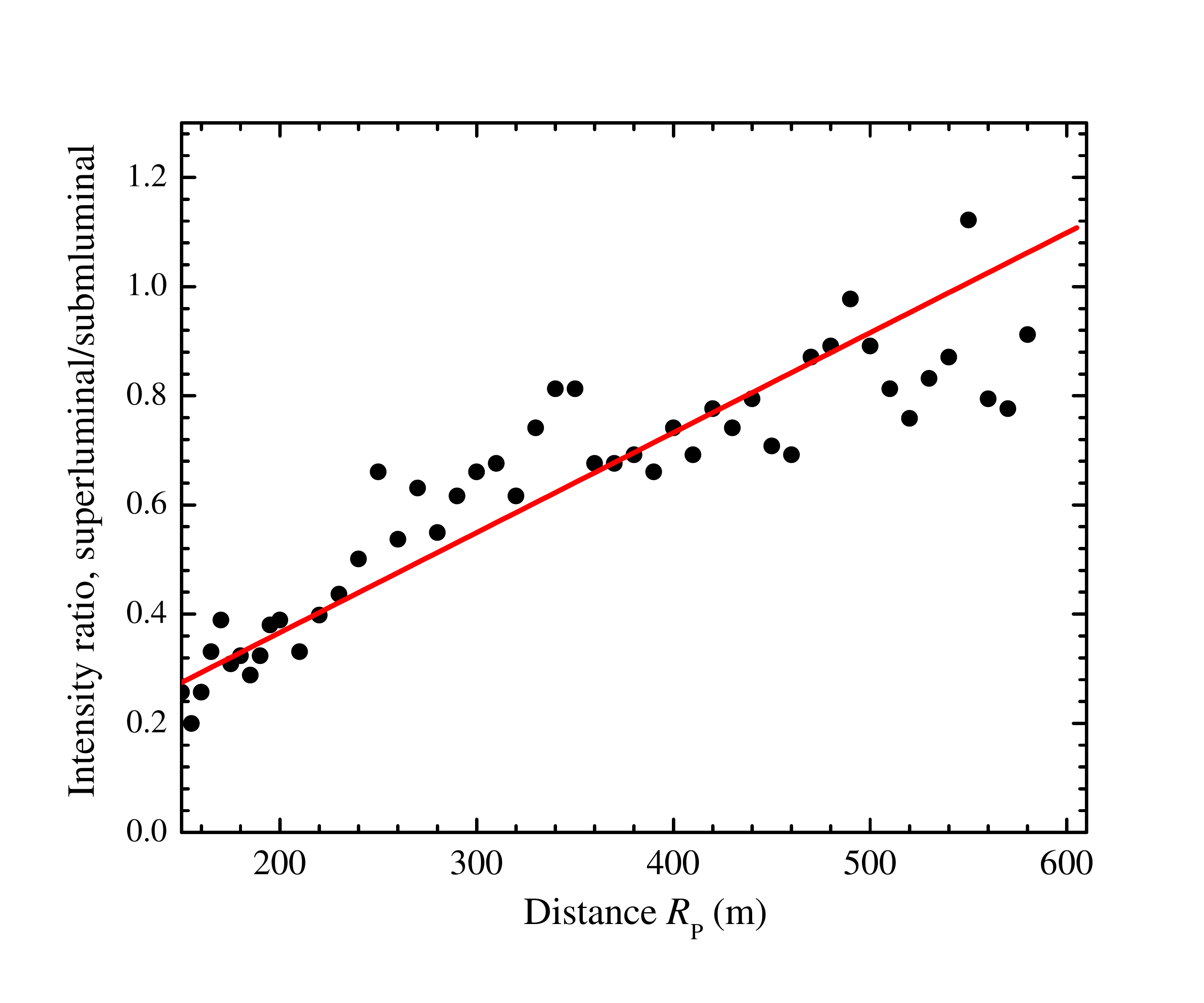}
\caption{(a) and (b): Ratio 
of the detected intensity with the experimental source (Fig.~\ref{fig0})
running
superluminally (at $v/c=1.064$) to that with the source running
subluminally ($v/c=0.875$) versus distance $R_{\rm P}$.
For both data sets the detector was moved along a path close to the
direction of the cusp expected for $v/c=1.064$. Data are points; the line is
a straight line through the origin, which implies that
the intensity in the superluminal case would decline in free
space as $1/R_{\rm P}$ (after Ref.~(6)).}
\label{exp}
\end{figure}

\begin{figure}
\centering
\includegraphics[angle=0,width=12.0cm]{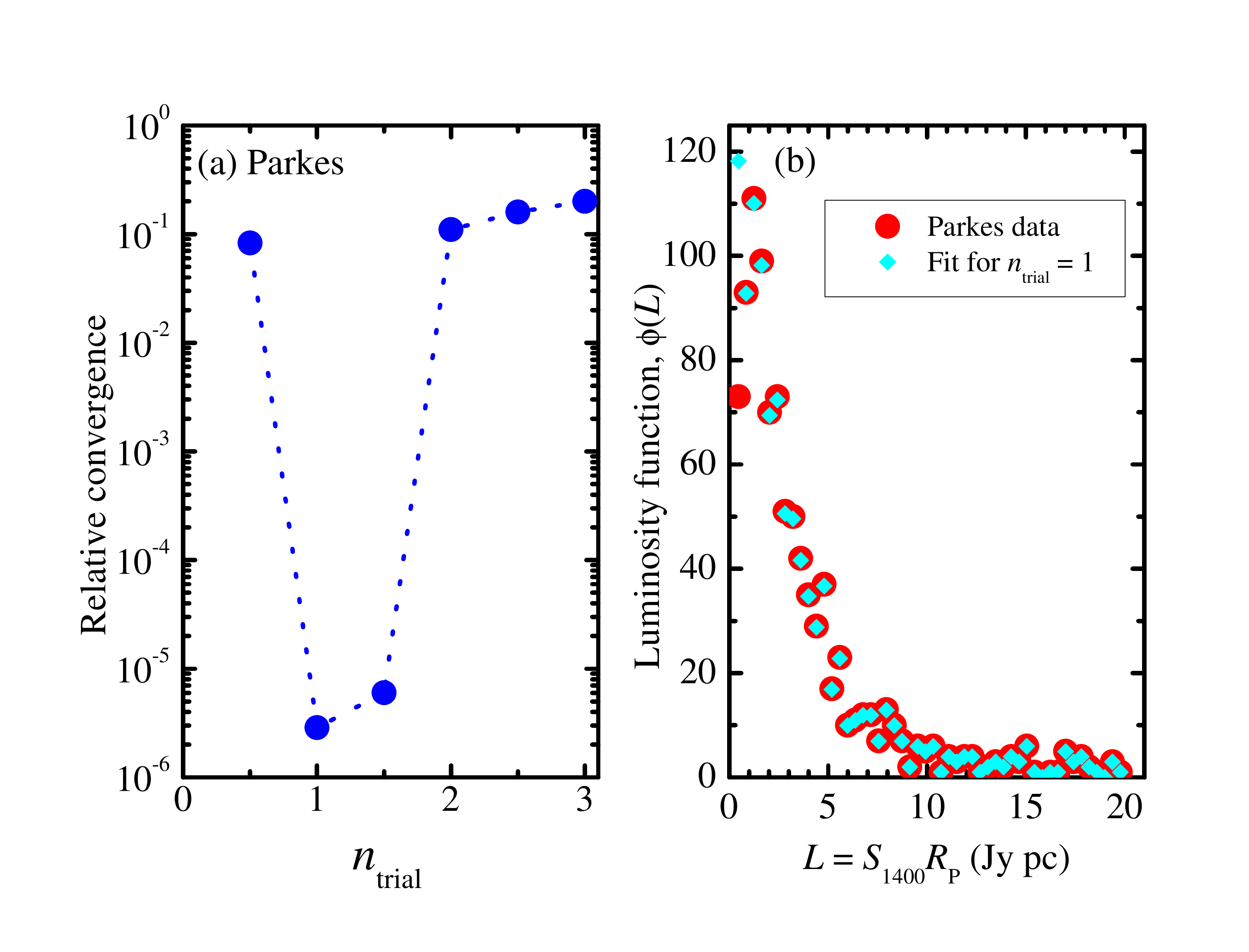}
\caption{Demonstration that the 
far-field emission of pulsars is dominated by
a component that decays nonspherically with
distance (after Ref.~(33)). 
(a)~The relative convergence error from 
the Maximum Likelihood Method
applied to 971 intragalactic pulsars (from the Parkes Multibeam Survey) 
as a function of the
trial exponent $n_{\rm trial}$. There is a clear minimum
at $n_{\rm trial}= 1$, strongly suggesting that
the flux at 1400~MHz, $S_{1400} \propto 1/R_{\rm P}$.
(b)~The inferred luminosity function
with raw pulsar data (taken from the Parkes Multibeam Survey) 
plotted as large circles,
and the fitted luminosity function as small diamonds. It is clear that
there is a very close correspondence between fit and data.}
\label{astFig2}
\end{figure}

Fig.~\ref{exp} shows a ground-based demonstration~\cite{ArdavanA:2004} using
the machine depicted in Fig.~\ref{fig0}. The
experiment consisted of moving a detector along the expected
cusp direction for a source speed of
$v/c = 1.064$ and recording the intensity as a function
of distance. The source speed was then set to
a subluminal velocity ({\it i.e.} no cusp formation- intensity follows
the inverse-square law with distance)
and the experiment repeated with the detector following the same path.
The ratio of the detected intensity for $v/c=1.064$
to that for the machine running at
subluminal speed increases
linearly with distance, showing that the cusp 
radiation intensity is decaying more slowly
with increasing distance, in line with expectations~\cite{ArdavanA:2004}.

For demonstrations covering much larger distances, one must
turn to astronomical observations.
In a recent paper~\cite{singleton}, the Maximum Likelihood Method~\cite{george} 
has been applied to observational data from 971 pulsars
to deduce the likely power law with which their flux
$S$ (a quantity proportional to intensity) falls off with distance. 
The Maximum Likelihood Method
determines the probable luminosity function
({\it i.e.} the distribution of pulsars with respect
to their radiated power) whilst making allowances
for missing observations due to the low sensitivity
of Earth-bound instruments~\cite{singleton}. The statistical success
of the method is measured using a relative convergence
error; to find the exponent $n$ in the relationship
$S\propto 1/R_{\rm P}^n$, the calculation is repeated for
several different values of a trial exponent, $n_{\rm trial}$ until
the value that gives the smallest relative convergence error
is located.  As can be seen in Fig.~\ref{astFig2},
the pulsar data are most convincingly fitted
by a flux $S\propto 1/R_{\rm P}$~\cite{singleton}.

In summary, experimental and observational data
suggest very strongly that the boundary
term neglected in Eq.~(2) dominates in the far field,
resulting in a component of the radiation that
is nonspherically decaying, with an intensity
proportional to $1/R_{\rm P}$. 

\section{Fundamental r\^ole of the retarded potential\label{sec:potential}}

While the results of an analysis
based on the retarded solution to the wave equation governing the field
depend crucially on the boundary conditions satisfied by the field at infinity
\cite{ArdavanH:2008a},
if we instead base our analysis
on the retarded potential,
we require no corresponding explicit knowledge
of the value of the potential in the radiation zone.
By using the field derived from the retarded potential [\Eq{8}],
we will evaluate the boundary term
in the retarded solution to the wave equation governing the field [\Eq{9}]
and show that,
far from being equal to zero,
as assumed in \Ref[s.]{Hannay:1996, Hannay:2000,Hannay:2001,Hannay:2006,Hannay:2008,Jackson:1999},
this boundary term constitutes the dominant contribution
to the value of the field in the far zone.
There is no discrepancy
between the results obtained from the retarded solution for the potential
and the retarded solution for the field
once the boundary term in the solution to the wave equation governing the field is retained.
Furthermore,
this analysis is robust
with respect to choice of integration boundaries (see the Appendix).

We first describe how the boundary contribution to the retarded
solution for the {\em potential} can always be made equal to zero,
irrespective of the source motion.

In the Lorenz gauge,
[\ie the choice of a set of potentials $(\vec A,A^0)$ that satisfy the Lorenz condition
$\vecsym{\nabla\cdot}\vec{A}+c^{-2}\partial A^0/\partial t=0$],
the electromagnetic fields
\begin{equation}
\vec{E}=-\nabla\subP A^0-\partial\vec{A}/\partial(c t\subP),
\qquad\vec{B}=\vecsym{\nabla}\subP\vecsym{\times}\vec{A},
\label{eq:3}
\end{equation}
are given by a four-potential $A^\mu$
that satisfies the wave equation
\begin{equation}
\nabla^2A^\mu-{1\over c^2}{\partial^2A^\mu\over\partial t^2}=
-{4\pi\over c}j^\mu,\qquad\mu=0,\cdots, 3,
\label{eq:4}
\end{equation}
where $A^0/c$ and $j^0/c$ are the electric potential and charge density
and $A^\mu$ and $j^\mu$ for $\mu=1,2,3$
are the Cartesian components of the magnetic potential $\vec{A}$
and the current density $\vec{j}$ \cite{Jackson:1999}.
The solution to the initial-boundary value problem for \Eq{4}
is given by
\begin{equation}\begin{split}
A^\mu(\vec{x}\subP,t\subP)=
&{1\over c}\int_0^{t\subP}{\rm d}t\int_V{\rm d}^3x\,j^\mu G
+{1\over4\pi}\int_0^{t\subP}{\rm d}t
\int_\Sigma{\rm d}\vec{S}\cdot(G\nabla A^\mu-A^\mu\nabla G)\\
&-{1\over 4\pi c^2}\int_V{\rm d}^3x
\Big(A^\mu{\partial G\over\partial t}-G{\partial A^\mu\over\partial t}\Big)_{t=0},
\end{split}\label{eq:5}
\end{equation}
in which $G$ is the Green's function
and $\Sigma$ is the surface enclosing the volume $V$
(see, \eg page 893 of \Ref{Morse:1953}).

The potential arising
from a general time-dependent localized source in unbounded space
decays as $R\subP^{-1}$ when $R\subP\equiv\vert\vec{x}\subP\vert\to\infty$,
so that for an arbitrary free-space potential
the second term in \Eq{5}
would be of the same order of magnitude
($\sim R\subP^{-1}$)
as the first term
in the limit that the boundary $\Sigma$ tends to infinity.
However,
even those potentials that satisfy the Lorenz condition
are arbitrary to within a solution of the homogeneous wave equation:
the gauge transformation
\begin{equation}
\vec{A}\to\vec{A}+\nabla\Lambda,
\qquad A^0\to A^0-\partial\Lambda/\partial t
\label{eq:6}
\end{equation}
preserves the Lorenz condition
if $\nabla^2\Lambda-c^{-2}\partial^2\Lambda/\partial t^2=0$
\cite{Besarab:2004}.
One can always use this gauge freedom in the choice of the potential
to render the boundary contribution
(the second term)
in \Eq{5}
equal to zero,
since this term, too,
satisfies the homogeneous wave equation.
Under the null initial conditions
$A^\mu|_{t=0}=(\partial A^\mu/\partial t)_{t=0}=0$
assumed in this paper,
the contribution from the third term in \Eq{5}
is identically zero.

In the absence of boundaries,
the retarded Green's function has the form
\begin{equation}
G(\vec{x}, t;\vec{x}\subP, t\subP)={\delta(t\subP-t-R/c)\over R},
\label{eq:7}
\end{equation}
where $\delta$ is the Dirac delta function
and $R$ is the magnitude of the separation
$\vec{R}\equiv\vec{x}\subP-\vec{x}$
between the observation point $\vec{x}\subP$
and the source point $\vec{x}$.
Irrespective of whether the radiation decays spherically
(as in the case of a conventional source) or nonspherically
(as applies for a rotating superluminal source---see \Sec{source})
\cite{ArdavanH:2007,ArdavanH:2008a},
therefore,
the potential $A^\mu$ due to a localized source distribution
that is switched on at $t=0$ in an unbounded space,
can be calculated from the first term in \Eq{5}:
\begin{equation}
A^\mu(\vec{x}\subP,t\subP)=
c^{-1}\int{\rm d}^3 x{\rm d}t\, j^\mu(\vec{x},t)\delta(t\subP-t-R/c)/R,
\label{eq:8}
\end{equation}
\ie from the classical expression for the retarded potential.
Whatever the Green's function for the problem may be,
in the presence of boundaries,
it approaches the expression in \Eq{7}
in the limit where the boundaries tend to infinity.
Hence one can also use this potential to calculate the field
on a boundary that lies at large distances from the source.

\section{Retarded solution of the equation governing the field\label{sec:field}}

We now return to the case of the {\em field} and show that an analogous
assumption about the boundary contribution may {\em not} be made.
Consider the wave equation
\begin{equation}
\nabla^2\vec{B}-{1\over c^2}{\partial^2\vec{B}\over\partial t^2}=
-{4\pi\over c}\vecsym{\nabla\times}\vec{j}
\label{eq:1}
\end{equation}
governing the magnetic field;
\Equation{1} may be obtained
by simply taking the curl of the wave equation for the vector potential
[\Eq{4} for $\mu=1,2,3$].
We write the solution to the initial-boundary value problem for \Eq{1},
in analogy with \Eq{5},
as
\begin{equation}\begin{split}
B_k(\vec{x}\subP,t\subP)=
&{1\over c}\int_0^{t\subP}{\rm d}t\int_V{\rm d}^3x\,(\vecsym{\nabla\times}\vec{j})_k G
+{1\over4\pi}\int_0^{t\subP}
{\rm d}t\int_\Sigma{\rm d}\vec{S}\cdot(G\nabla B_k-B_k\nabla G)\\
&-{1\over 4\pi c^2}\int_V{\rm d}^3x
\Big(
B_k{\partial G\over\partial t}-G{\partial B_k\over\partial t}
\Big)_{t=0},
\end{split}\label{eq:9}
\end{equation}
where $k=1,2,3$ designate the Cartesian components
of $\vec{B}$ and $\vecsym{\nabla\times}\vec{j}$.

Here, we no longer have the freedom,
offered by a gauge transformation in the case of \Eq{5},
to make the boundary term zero,
nor does this term always decay faster than the source term,
so that it could be neglected for a boundary that tends to infinity,
as is commonly assumed in textbooks
(\eg page 246 of \Ref{Jackson:1999})
and the published literature \cite{Hannay:1996,Hannay:2000,Hannay:2001,Hannay:2006,Hannay:2008}.
The boundary contribution
to the retarded solution of the wave equation governing the field
[the second term on the right-hand side of \Eq{9}]
entails a surface integral
over the boundary values of both the field and its gradient.
For the rotating superluminal source (\ref{eq:-1}),
where the gradient of the field increases
as $R\subP^{7/2}$
over a solid angle that decreases as $R\subP^{-4}$,
this boundary contribution
is proportional to $R\subP^{-1/2}$
(see \Ref{ArdavanH:2008a}).
Not only is this not negligible
relative to the contribution from the source term
[the first term on the right-hand side of \Eq{9}],
{\em but the boundary term constitutes the dominant contribution
to the radiation field in this case} \cite{ArdavanH:2008a}.

If one ignores the boundary term
in the retarded solution of the wave equation governing the field,
as in \Ref[s.]{Hannay:1996,Hannay:2000,Hannay:2001,Hannay:2006,Hannay:2008,Jackson:1999},
one obtains a different result,
in the superluminal regime,
from that obtained
by calculating the field via the retarded potential
\cite{ArdavanH:2007,ArdavanH:2006,ArdavanH:2008b}.
This apparent contradiction
stems solely from having ignored a term
in the solution to the wave equation
that is, by a factor of the order of $R\subP^{1/2}$,
larger than the term that is normally kept.
The contradiction disappears
once we take the neglected term into account:
the solutions to the wave equations
governing both the potential and the field
predict that the field of a rotating superluminal source
decays as $R\subP^{-1/2}$
as $R\subP$ tends to infinity,
a result that has been demonstrated experimentally
\cite{ArdavanA:2004,Singleton:2004}.

We note, furthermore, that the representation
\begin{equation}
\vec{A}(\vec{x}\subP,t\subP)=\frac{1}{c}\int{\rm d}^3x\,
\frac{[\vec{j}(\vec{x},t)]}{\vert\vec{x}-\vec{x}\subP\vert},
\label{eq:12}
\end{equation}
of the retarded potential is differentiable
as a classical (as opposed to generalized) function
only in the case of a moving source whose speed
does not exceed that of the waves it generates.
The steps, familiar from the subluminal regime,
one takes to derive \Eq{2}
by differentiating \Eq{12} as a classical function
are not mathematically permissible
when the moving source has volume elements
that approach the observer with the wave speed and zero acceleration at the retarded time.
\cite{ArdavanH:1999,ArdavanH:2000}
Contrary to the usual assumption
\cite{Hannay:1996,Hannay:2000,Hannay:2001,Hannay:2006,Hannay:2008},
the retarded distribution of the density of a moving source
is not necessarily smooth and differentiable if its rest-frame distribution is.
The retarded distribution of a rotating source with a moderate superluminal speed
is in general spread over three disjoint volumes
(differing in shape from each other
and from the volume occupied by the source in its rest frame)
whose boundaries depend on the spacetime position of the observer.
The limits of integration in \Eq{12},
which delineate the boundaries of the retarded distribution of a localized source,
are not differentiable functions of the coordinates of the observer
at those points on the source boundary
that approach the observer, along the radiation direction,
with the speed of light at the retarded time \cite{ArdavanH:1999,ArdavanH:2000}.
In the superluminal regime,
derivatives of the integral representing the retarded potential
are well-defined only as generalized functions \cite{ArdavanH:2004}.

\section{Summary and conclusions\label{sec:summary}}
We must stress
that we have not calculated the radiation field
directly from the solution to the wave equation governing the field
(the solution whose boundary term is normally neglected 
\cite{Hannay:1996,Hannay:2000,Hannay:2001,Hannay:2006,Hannay:2008}).
We have instead first solved the wave equation governing the potential
(whose solution has no boundary term)
and then used this solution
to evaluate the neglected term
in the exact version of the retarded solution for the field [\Eq{9}].
What one obtains by including the boundary term
in the retarded solution to the wave equation governing the field
is merely a mathematical identity;
it is not a solution
that could be used to calculate the field
arising from a given source distribution in free space.
Unless its boundary term
happens to be small enough relative to its source term to be neglected,
a condition that cannot be known {\em a priori},
the solution in question would require
that one prescribe the field in the radiation zone
(\ie what one is seeking)
as a boundary condition.
Thus,
the role of the classical expression
for the retarded potential in radiation theory
is much more fundamental
than that of the corresponding retarded solution
of the wave equation governing the field.
The only way to calculate the free-space radiation field
of an accelerated superluminal source
is to calculate the retarded potential
and differentiate the resulting expression
to find the field \cite{ArdavanH:2007,ArdavanH:2008a}
(see also \Ref[s.]{ArdavanH:2006,ArdavanH:2008b}).

\vskip1truecm
A.\ A.\ is supported by the Royal Society.
J.\ S., J.\ F., and A.\ S.\ are supported by U.S.\ Department of Energy grant LDRD 20080085DR,
``Construction and use of superluminal emission technology demonstrators
with applications in radar, astrophysics and secure communications.''

\renewcommand{\refname}{References and notes}

\appendix
\section{A note on integration boundaries\label{sec:boundaries}}

It has been suggested \cite{Hannay:2008}
that our demonstration of the nonspherical decay
of the field of a rotating superluminal source
can be dismissed
on the grounds that the boundary term in \Eq{9} must be identically zero.
However,
the claim that the boundary contribution
to the retarded solution of the wave equation governing the field
is exactly zero even when the boundary encloses a source
violates the very foundations of diffraction theory:
For a volume $V$ in which no sources are present,
\Eq{9} reduces to
\begin{equation}
B_k({\bf x}\subP,t\subP)=
{1\over4\pi}\int_0^{t\subP}
{\rm d}t\int_\Sigma{\rm d}{\bf S}\cdot(G\nabla B_k-B_k\nabla G),
\label{eq:10}
\end{equation}
under the null initial conditions $B_k|_{t=0}=(\partial B_k/\partial t)_{t=0}=0$.
As in the customary geometry for diffraction,
the closed surface $\Sigma$ can consist of two disjoint closed surfaces,
$\Sigma_{\rm inner}$ and $\Sigma_{\rm outer}$
(\eg two concentric spheres),
both of which enclose the source
(see Fig.~10.7 of \Ref{Jackson:1999}).
If the observation point does not lie
in the region between $\Sigma_{\rm inner}$ and $\Sigma_{\rm outer}$,
\ie lies outside the closed surface $\Sigma$,
then the composite surface integral in \Eq{10} vanishes:
\begin{equation}\begin{split}
\int_0^{t\subP}
{\rm d}t\int_\Sigma{\rm d}{\bf S}\cdot(G\nabla B_k-B_k\nabla G)
&=\int_0^{t\subP}
{\rm d}t\left(\int_{\Sigma_{\rm inner}}+\int_{\Sigma_{\rm outer}}\right){\rm d}{\bf S}\cdot(G\nabla B_k-B_k\nabla G)\\
&=0
\end{split}\label{eq:11}
\end{equation}
(see \Ref{Morse:1953}).
Under no circumstances, however,
would the integrals over $\Sigma_{\rm inner}$ or $\Sigma_{\rm outer}$
vanish individually if these surfaces enclose a source,
\ie if there is a nonzero field inside $\Sigma_{\rm inner}$;
nor does the invariance of the values of these integrals
under deformations of $\Sigma_{\rm inner}$ and $\Sigma_{\rm outer}$
have any bearing on whether they are nonvanishing or not \cite{Hannay:2008}.
\Equation{10} forms the basis of diffraction theory \cite{Jackson:1999}.
If the surface integrals over $\Sigma_{\rm inner}$
and $\Sigma_{\rm outer}$ were to vanish individually,
as claimed in \Ref{Hannay:2008},
the diffraction of electromagnetic waves
through apertures on a surface enclosing a source
would be impossible
(see Sec.~10.5 of \Ref{Jackson:1999}).


\begin{thebibliography}{99}
\bibitem{note:conservation}
Conservation of energy is not violated
by the subbeams; enhancement of the radiation fields
within the subbeams is exactly compensated by reduction
of the polar widths of the subbeams, so that the flux of energy across their cross sections remains the same at all distances from the source. The physical mechanism underlying the
properties of the subbeams is the focusing in the time domain
possible when a source both travels faster than its own
waves and accelerates;
contributions towards the intensity of
the radiation made over an extended period of emission time
are received during a significantly shorter period of observation time.
\end{thebibliography}
\end{document}